# HST polarization map of the ultraviolet emission from the outer jet in M87 and a comparison with the 2cm radio emission.

R. C. Thomson[1], D. R. T. Robinson[1], N. R. Tanvir[1], C. D. Mackay[1] and A. Boksenberg[2]

1. *Institute of Astronomy, Madingley Road, Cambridge CB3 0HA.*

2. *Royal Greenwich Observatory, Madingley Road, Cambridge CB3 0EZ.*

## ABSTRACT

We present the first high resolution polarization map of the ultraviolet emission from the outer jet in M87. The data were obtained by the Faint Object Camera (*FOC*) on the Hubble Space Telescope. The polarization map has a resolution of 0.2 arcsec and was derived using data from three linearly polarized images combined with the best available calibration data. The ultraviolet emission is highly polarized ($\sim 40\%$) with the magnetic vector aligned roughly with the jet axis, except in the region of the brightest knot (Knot A) where the position angle changes abruptly and the magnetic vector becomes perpendicular to the jet axis. A similar behaviour is seen in the 2cm *VLA* radio polarization map. By aligning the *FOC* and *VLA* data, we present ultraviolet–2cm spectral index, depolarization and rotation measure maps. We identify a region of high depolarization adjacent to Knot A. This is the first direct observational evidence that indicates the presence of a cloud or filament of dense thermal material which is mixed with the synchrotron emitting plasma of the jet. The interaction of the jet with this cloud is likely to be responsible for the sudden increase in the brightness of the jet at Knot A due to an induced shock. We suggest that the dark line seen in the 2cm radio data between Knot A and Knot C could be the shadow or magnetotail of the depolarizing cloud in the jet.

## 1 INTRODUCTION

M87 is a massive elliptical galaxy near the centre of the Virgo cluster with an optical jet that extends 20 arcsec from the nucleus. At an assumed distance of 15 Mpc, where 1 arcsec corresponds to 75 pc, the jet is 1.5 kpc in projection — well within the effective radius of the galaxy. The jet in M87 is well resolved at radio wavelengths (Owen, Hardee & Cornwell 1989). The radio emission from the inner jet (between the nucleus and the brightest knot, Knot A) appears to be edge-brightened and dominated by helical structure. Both the radio emission and the optical emission suddenly increase in brightness at Knot A where the jet also broadens before merging with the 2 kpc radio lobes beyond Knot C.

Ground-based optical images of the jet are hampered by the seeing, although a resolution of 0.25 arcsec (FWHM) was claimed by Stiavelli *et al.* (1992) by deconvolving 0.6 arcsec ground-based images. The ground-based data is also limited to optical wavelengths where the emission from the jet is superimposed on the underlying galaxy profile. M87 looks very different at ultraviolet wavelengths where the underlying galaxy is hardly detectable. Ultraviolet images of the jet in M87 have been published by Boksenberg *et al.* (1992). These were obtained using the *FOC f/96* camera on the *HST*. The deconvolved *FOC* images of the jet show a striking similarity to the highest resolution radio data, although there are important differences (Macchetto 1992).

In this paper, we present the polarization data obtained by the *FOC f/96* camera with the F220W filter. Although polarized images were also obtained using the F430W (B band) filter (where the underlying galaxy is bright), the telescope was moved between images with only a small overlap region. We therefore chose to work with the ultraviolet (F220W) images only. These images cover the brightest part of the jet between 10–20 arcsec from the nucleus, including Knots A, B and C.

In Section 2 we present the polarized ultraviolet images and detail our data reduction methods, including deconvolution techniques and calculation of the ultraviolet polarization map. In Section 3, we compare our results with the 2cm *VLA* radio data of similar resolution, and discuss these results in terms of contemporary models of the jet kinematics and structure in Section 4.

## 2 OBSERVATIONS AND DATA REDUCTION

Fig. 1 shows the three polarized images of the jet in M87 obtained on 17 April 1991 by the *FOC f/96* camera (512×512 pixel$^2$ format; pixel size 0.022×0.022 arcsec$^2$) on the *HST*. The images were taken through a broad band ultraviolet filter (F220W) in combination with the three polarizers POL0, POL60 and POL120 consecutively. Each image has the same exposure time of 1496 s and has been flat fielded and geometrically corrected in the standard way using the calibration data available at that time. In addition the reseau marks have been removed using the STSDAS task *rremovex* in the *FOC* package. The derived total intensity image is also shown in Fig. 1. This was obtained by summing the three polarized images after alignment and weighting by the inverse polarizer throughputs (see Section 2.2).

These images have limited S/N with peak values of 25, 40 and 33 count pixel$^{-1}$ and mean background values of 0.84, 0.75 and 0.78 count pixel$^{-1}$ in the POL0, POL60 and POL120 images respectively. The typical peak count rates are therefore well within the linearity limit of the *FOC f/96* photon counting detector (∼ 0.1 count pixel$^{-1}$ s$^{-1}$ for extended emission). Even with such limited S/N, the effect of the spherical aberration can be seen as a halo around the jet. For this reason it was necessary to deconvolve the *FOC* images before comparison with the radio data.

### 2.1 DECONVOLUTION PROCEDURE

We used a Maximum Entropy method (MEMSYS-5, Gull & Skilling 1991) to deconvolve the data. This is the same method we had used previously on *FOC* images with point sources (Thomson *et al.* 1993). Before deconvolution, a realistic point spread function (PSF) had to be prepared. The resulting quality of the deconvolved image depends on the accuracy and S/N of both the image and the PSF. At the focus position relevant to these observations, the *FOC* PSF library contains point source images taken through medium band filters only. We therefore decided to use a hybrid PSF made up from 4 medium band filter PSFs, namely: F190M, F210M, F231M and F253M. The individual PSF images were background subtracted and aligned with each other to an estimated accuracy of 0.1 pixel. The aligned images were then added together with weighting factors given by the published transmission curve of the F220W filter (Paresce 1991) and determined at the wavelength of peak transmission for the corresponding medium band filter. Finally, all pixel values outside the central 256×256 pixel$^2$ were set to zero in order to minimise noise since the PSF is confined within this region.

The resulting PSF has a FWHM=4.7 pixels and a central pixel value of 1436 counts. We used the same PSF for all three polarized images. In fact, the polarizers have slightly different PSFs with the aberrated optics, but this distinction would be necessary only with data of much higher S/N than we are dealing with here.

The polarized images were prepared for deconvolution by replacing the aberrant border pixels values (mostly zero), introduced during the geometrical correction procedure, with a more suitable constant value set to the mean background value measured from the image. This is done in order to avoid problems caused by the artificial sharp edges that do not have the imprint of the instrument PSF.

Since these images have limited S/N and contain extended emission, care must be taken to choose an appropriate value for the effective resolution of the reconstructed image. MEMSYS-5 allows the user to impose an intrinsic pixel-pixel correlation on the reconstucted image, which we refer to as the intrinsic correlation function (ICF). Without such a facility, the deconvolution of noisy data to single pixel resolution becomes dominated by peaks in the noise. In such cases the reconstructed image must be smoothed to give a reliable reconstruction from the data.

We used a gaussian ICF (as implemented in MEMSYS-5), and determined the best value for the smoothing length by deconvolving one of the polarized images (POL0) using different values for the standard deviation of the gaussian ICF. We chose a value of 4 pixels for the standard deviation (corresponding to a resolution of 9 pixels or 0.2 arcsec FWHM) by requiring the deconvolved image to be the sharpest representation of the raw data without introducing any significant noise generated structure.

Having determined a reasonable value for the ICF smoothing length, the polarized images were deconvolved using MEMSYS-5, including background fitting. This effectively removes much of the noisy background which otherwise makes the reconstructed image appear mottled. The reconstructed polarized images and corresponding total intensity image are shown in Fig. 2

### 2.2 POLARIZATION DETERMINATION

Before calculating the polarization map, the deconvolved polarized images were shifted to compensate for the unequal offsets in the image position introduced by the three polarizers. The absolute offsets (relative to positions without the polarizers) have been measured by Hodge (1993) and are given in Table 1. Note the significantly different offset introduced by the polarizer POL60, which is also characterized by a poor throughput shortward of 2200Å that must be compensated for when calculating the ultraviolet polarization.

The signal obtained through polarizer $i$ which passes linearly polarized light at position angle $\theta_i$ is a function of the Stokes parameters ($I, Q$ and $U$) of the incoming beam (Lupie 1984):

$$S_i = \frac{1}{2} g_i (I + k_i Q \cos 2\theta_i + k_i U \sin 2\theta_i).$$

The polarizer angles ($\theta_i$), thoughputs ($g_i$) and polarization efficiencies ($k_i$) used for this work are given in Table 1. We estimated the relative throughput of POL60 when used with the ultraviolet filter F220W by calculating the combined throughput (filter + optical telescope assembly + FOC detector quantum efficiency + polarizer) for each polarizer using the available wavelength calibration files. We then integrated the combined throughput over the appropriate wavelength range. The results for POL0 and POL120 were identical to within 1% (as expected), and we found the relative throughput of POL60 to be 0.75, as given in Table 1.

Defining the corrected intensities

$$s_i = \frac{S_i}{g_i},$$

then (with $k_1 = k_2 = k_3 = 1$) the normalized Stokes parameters (in instrumental units)

$$I = \frac{2(s_1 + s_2 + s_3)}{3},$$

$$q = \frac{Q}{I} = \frac{2s_1 - s_2 - s_3}{s_1 + s_2 + s_3},$$

and

$$u = \frac{U}{I} = \frac{\sqrt{3}(s_2 - s_3)}{s_1 + s_2 + s_3}.$$

The fractional polarization is then given by

$$p = (q^2 + u^2)^{\frac{1}{2}},$$

and the polarization position angle,

$$\theta = \frac{1}{2} \arctan \frac{u}{q},$$

which is measured anticlockwise from the scan direction ($x$-axis). Note that no absolute flux calibration is required to make accurate polarization measurements, only the relative intensities are required.

Fig. 3 shows the polarization map derived from the deconvolved polarized images superimposed on a countour plot of the reconstructed total intensity image. The polarization map shows the orientation of the *magnetic* vectors which, for synchrotron radiation, is parallel to the direction of the resolved magnetic field component projected onto the plane of the sky. To obtain this polarization map, the shifted images were reduced from 512×512 to 64×64 by summing over 8×8 pixels (which is marginally less than the effective resolution of the reconstructed images) to improve the S/N. We estimate the *relative* errors in the fractional polarization to be $\lesssim$20% with a corresponding position angle error of $\lesssim 5^0$. There may be significant systematic errors due to the imperfect calibration ($\Delta x, \Delta y, \theta_i, g_i$ and $k_i$), but we think that these are smaller than our estimated errors.

# 3 COMPARISON WITH THE RADIO EMISSION

We compare the ultraviolet emission and polarization as measured by the *FOC* with the 2cm radio emission and polarization as measured by the *VLA* (Owen, Hardee & Cornwell 1989). After alignment, we present the ultraviolet–2cm spectral index map, the ultraviolet–2cm depolarization map, and the ultraviolet–2cm rotation measure map.

## 3.1 Alignment of the FOC and VLA data sets

The exact location of the *FOC* image on the sky is not known with sufficient accuracy from the *HST* engineering data for comparison with the *VLA* data. The *FOC* image also does not contain any point source (the nucleus of M87 for example) which can be used for accurate astrometry. Given the striking similarity of the ultraviolet and radio emission (Boksenberg *et al.* 1992), we chose to align the *FOC* and *VLA* data sets by cross-correlation of the total intensity maps. As such, any systematic offset of the ultraviolet and radio data will not be detected, however, we think this is unlikely to be the case.

Due to the lower resolution (0.2 arcsec) of the *FOC* data compared to the 2cm *VLA* radio data (0.14 arcsec), the radio map was first smoothed to the same resolution using the appropriate gaussian filter. The total intensity *FOC* image was then re-sampled on a coarser pixel grid to give the same pixel size as the 2cm radio data (0.045 arcsec). We then used an iterative procedure to properly register the *FOC* and *VLA* data. By cross-correlating the total intensity images we first determined the linear offset of the two data sets. We then repeated the cross-correlation for various position angles of the *FOC* data relative to the *VLA* data to obtain the best fit position angle and linear offset independent of the *HST* engineering data. By this means we were able to align the *FOC* and *VLA* data to better than 0.05 arcsec. Having properly registered the total intensity *FOC* image with the total intensity radio map, we then aligned the individual polarized *FOC* images with the radio data using the transformation derived from the total intensity image and recalculated the ultraviolet polarization map in the new reference frame.

## 3.2 The ultraviolet–2cm spectral index map

Fig. 4 shows the ultraviolet–2cm spectral index map. The spectral index, $\alpha$ ($S_\nu \propto \nu^{-\alpha}$), was calculated as the logarithm of the flux ratios. As can be seen, the spectral index is remarkably uniform over the main body of the jet. This is just another way of saying that the ultraviolet and radio emission are very similar, as described by Boksenberg *et al.* (1992). On the south side of the jet between Knot A and Knot B, however, the radio emission extends somewhat further than the ultraviolet emission and the spectral index steepens accordingly. This has also been noted by Macchetto (1992) and Biretta (1993) who found evidence for a more subtle extension of the radio emission throughout the jet using higher resolution unpolarized FOC images.

## 3.3 The ultraviolet–2cm depolarization map

Fig. 5 shows the ultraviolet–2cm depolarization map. The ultraviolet–2cm depolarization is defined as the ratio of the fractional polarization as observed by the *FOC* in the ultraviolet divided by the fractional polarization observed by the *VLA* at 2cm, after smoothing to the same resolution as the *FOC* data. The depolarization over most of the jet is small (ie, $\sim 1$), with one significant exception. A region of large depolarization is clearly detected along the NW edge of Knot A. This cannot be due to the flatter spectral index of the radio emission (0.5 compared to 1 for the ultraviolet emission) which would reduce the fractional polarization by only 10%. Such strong depolarization indicates the presence of Faraday active thermal plasma mixed with the synchrotron emitting plasma at this location (Burn 1966).

## 3.4 The ultraviolet–2cm rotation measure map

Fig. 6 shows the ultraviolet–2cm rotation measure map. The rotation measure

$$RM = 2500(\theta_{2cm} - \theta_{uv}) \text{ rad m}^{-2}.$$

is calculated using the position angles of the 2cm radio data ($\theta_{2cm}$) and the *FOC* ultraviolet data ($\theta_{uv}$). The random error of the *FOC* position angles are estimated to be $\sim 5^0$ which corresponds to an error $\sim 200$ rad m$^{-2}$ in the ultraviolet–2cm rotation measure map.

Large ($\gtrsim 1000$ rad m$^{-2}$) rotation measures are present in some parts of the jet, as reported by Owen, Eilek & Keel (1990). It is thought that most, if not all, of this Faraday rotation is produced in thermal material which forms a foreground screen, rather than being mixed with the synchrotron emitting plasma. Although necessarily noisy (ie, mottled), the ultraviolet–2cm rotation measure map shows an anomalous region near the NW edge of Knot A that coincides with the anomalous feature in the depolarization map. The association of these features with the sudden brightening of the jet at Knot A indicates the presence of a filament or cloud of Faraday active plasma mixed with the synchrotron emitting plasma of the jet at this location.

## 4 DISCUSSION

At ultraviolet wavelengths, Faraday rotation and depolarization are negligible, and the fractional polarization of (optically thin) synchrotron emission is determined by the spectral index, $\alpha$ ($S_\nu \propto \nu^{-\alpha}$), and the relative magnitudes of the unresolved (assumed random) magnetic field component, $B_{ran}$, and the resolved uniform component, $B_{uni}$ (Burn 1966)

$$p = p(\alpha) \frac{B_{uni}^2}{(B_{uni}^2 + B_{ran}^2)}$$

where

$$p(\alpha) = \frac{3\alpha + 3}{3\alpha + 5}.$$

For the ultraviolet emission, $\alpha \simeq 1$ (Boksenberg *et al.* 1992) and hence $p(\alpha) \simeq 0.75$. This is the maximum fractional polarization which could be observed from a region with a completely uniform magnetic field. The high fractional polarization observed in Knot A implies that the magnetic field is well resolved and uniform. The alignment of the magnetic field nearly perpendicular to the jet axis at Knot A is consistent with compression of the magnetic field in the jet at a shock front.

Minimum pressure arguments give a magnetic field density of $\sim 300\mu G$ in this region (referred to as $A_{34}$ in Owen, Hardee & Cornwell 1989) with a corresponding lifetime $\sim 15$ yr for the high energy particles that emit at ultraviolet wavelengths. These particles can travel only 5 pc (0.06 arcsec) from their site of origin. The 2cm radio emitting particles, however, have corresponding lifetimes $\sim 4,000$ yr and can travel over 1 kpc (13 arcsec) from their site of origin. It is therefore surprising to see that the radio emission from the jet has almost the same spatial extent as the ultraviolet emission. If the magnetic field were highly tangled, then neither the ultraviolet nor the radio emitting particles could travel far from the site of origin. However, the regular structure combined with the high polarization observed in Knot A, implies that the magnetic field must be nearly uniform. In order to explain the apparent similarity of the ultraviolet and 2cm radio emission, the emissivity must be constrained not by the particle lifetimes, but by the spatial extent of the magnetic field. One possible mechanism is that the longer lasting particles escape from the intense magnetic field within the jet into the surrounding ISM where the field intensity is much lower.

For synchrotron emission confined to the surface of a cylinder, as suggested by Owen, Hardee & Cornwell (1989), we would expect the fractional polarization to vary systematically across the cylinder, exhibiting maximum polarization at the edges, where even the random field component (which is confined to the surface layer) must be aligned with the jet. The 2cm radio data appear to be consistent with this interpretation for the inner jet, but in the outer jet, the fractional polarization appears to be highest in the centre of Knot A and not at the edge. The edge effect may be present in some parts of the ultraviolet map, but is not obvious.

The observation of proper motions in the inner jet of M87 (Biretta 1993) reveal features which appear to move with a significant fraction of the speed of light (between 0.1c and 0.6c). Two of the condensations in Knot D show superluminal velocities of 1.7c and 3.0c respectively. These kinematic data are consistent with a relativistic inner jet having a Lorentz factor $\gamma > 3$ and oriented $\sim 40^0$ with respect to the line of sight. If so, relativistic beaming may increase the brightness of the near-side inner jet by a factor $> 150$.

Biretta (1993) describes a kinematic model of the whole jet, in which the jet undergoes a shock at Knot A and continues to lose kinetic energy in the transition region between knots A and C until it finally merges with the 2kpc lobe outside Knot C. Our discovery of strong deloparization in the jet immediately NW of Knot A strongly suggests the presence of a dense thermal filament or cloud at this point. The presence of such a cloud is consistent with Biretta's kinematic description of the jet, providing a suitable obstacle to induce the shock at Knot A. The flip in the polarization position angle at Knot A where the projected magnetic field becomes perpendicular to the jet axis is consistent with the compression expected at a shock caused by the impact of the jet with such an obstacle.

We can estimate a lower limit for the mean electron density in the cloud by assuming a simple slab model (Burn 1966). Such a source will be significantly depolarized at a wavelength $\lambda_d$ (m) given by

$$\lambda_d^{-2} = n_e B_\parallel L,$$

where $n_e$ is the mean electron density (cm$^{-3}$), $B_\parallel$ is the mean field component parallel to the line of sight ($\mu$G), and $L$ is the path length (pc). For a mean field of 300 $\mu$G, and a path length of 75 pc (jet width), a mean electron density $\gtrsim 0.1$ cm$^{-3}$ is required to depolarize the jet at 2cm.

The sudden jump in emission from the jet at Knot A is presumably due to particle acceleration in a shock. Thus relativistic beaming is not the only mechanism which makes the jet in M87 appear to be one-sided — interaction with the local ISM also produces a significant increase in the jet brightness. The lack of emission from the putative counter-jet could be due to the lack of thermal filaments in the ISM on that side of the galaxy. Narrow band H$_\alpha$+[NII] images of M87 (Sparks, Ford & Kinney 1993) show many gaseous filaments on the jet side, but none on the counter jet side. The putative counter-jet therefore remains invisible due to the lack of any shock induced structure and relativistic beaming of the unperturbed jet.

Owen, Hardee & Cornwell (1989) reported the discovery of a dark line in the radio emission which runs down the interior of the jet between knots A and C. This dark line is a dip in the intensity at about the 15% level. It is a narrow feature which requires both high resolution and high dynamic range to detect. It is not seen in our FOC data, although this could be due to the lower resolution and dynamic range of these data. The dark line is also not seen in our spectral index map (Fig. 4), where the radio data have been smoothed to the same resolution as the FOC data. Interestingly, however, there is some evidence for such a feature in the depolarization and rotation measure maps (Figs. 5 and 6). Given our interpretaton of these data in terms of an interaction between the jet and a dense gas cloud, we suggest that the dark lane seen in the 2cm *VLA* radio data could be the shadow or magnetotail of the depolarizing cloud in the jet.

We have presented the first high resolution polarization map of the jet in M87 and compared our results with the 2cm *VLA* radio polarization data. A region of strong depolarization is identified adjacent to the brightest knot in the jet. We interpret these results in terms of Faraday depolarization by a dense gas cloud or filament in the interstellar medium of M87 with which the jet is interacting. Higher resolution and improved S/N data of the ultraviolet (or optical) emission using the refurbished FOC should reveal the structure of this interesting region when compared with the 2cm *VLA* data without the need for deconvolution.

## 5   ACKNOWLEDGEMENTS

We are grateful to F. N. Owen for kindly making available the 2cm *VLA* polarization data. We thank Richard Sword for help in preparing the figures.

## TABLE AND FIGURE CAPTIONS

**Table 1.**
Polarizer offsets ($\Delta x, \Delta y$) measured in pixels

(from Hodge 1993); position angles ($\theta_i$); transmission coeficients ($g_i$) and polarization efficiencies ($k_i$).

| Polarizer | $\Delta x$ | $\Delta y$ | $\theta_i$ | $g_i$ | $k_i$ |
|---|---|---|---|---|---|
| Pol0 | 1.4 | -7.4 | 0 | 1.0 | 1.0 |
| Pol60 | -2.3 | -9.2 | 60 | 0.75 | 1.0 |
| Pol120 | 1.4 | -6.6 | 120 | 1.0 | 1.0 |

**Figure 1.** Polarized *FOC* images of the jet in M87: a) F220W+POL0, b) F220W+POL60, c) F220W+POL120, and d) total intensity image. The location of knots A, B and C is also shown.

**Figure 2.** MaxEnt deconvolutions of the *FOC* images shown in Fig. 1. The effective resolution of the reconstructed images is 0.2 arcsec.

**Figure 3.** Ultraviolet polarization map of the jet in M87 superimposed on a contour plot of the reconstructed *FOC* image. The map shows the orientation of the *magnetic* field vectors.

**Figure 4.** Ultraviolet–2cm spectral index map. The linear grey scale calibration assumes that the optical–radio spectral index of Knot A is 0.62, as published by Biretta, Stern & Harris (1991).

**Figure 5.** Ultraviolet–2cm depolarization map. Regions of strong depolarization (ie, where the radio polarization is much smaller than the optical polarization) appear dark in this linear grey scale representation.

**Figure 6.** Ultraviolet—2cm rotation measure map. Positive rotation measures appear dark and negative rotation measures appear light in this linear grey scale representation. Where necessary, the *FOC* polarization position angles have been rotated by $\pm\pi$ rad to give the smallest possible magnitude for the rotation measure.